# Impact of geometry on chemical analysis exemplified for photoelectron spectroscopy of black silicon

*Jens U. Neurohr,*[‡] *Anton Wittig,*[‡] *Hendrik Hähl,*[‡] *Friederike Nolle,*[‡,∥] *Thomas Faidt,*[‡] *Samuel Grandthyll,*[‡] *Karin Jacobs,*[‡] *Michael A. Klatt,*[*§] *and Frank Müller*[†,‡]

[‡]*Experimental Physics and Center for Biophysics, Saarland University, Campus E2 9, 66123 Saarbrücken, Germany*

[§]*German Aerospace Center (DLR), Institute for AI Safety and Security, Wilhelm-Runge-Str. 10, 89081 Ulm, Germany; German Aerospace Center (DLR), Institute for Material Physics in Space, 51170 Köln, Germany; Department of Physics, Ludwig-Maximilians-Universität München, Schellingstr. 4, 80799 Munich, Germany*

[∥] *Department of Electrical Engineering, Trier University of Applied Science, Schneidershof, Trier, Germany*

[*] michael.klatt@dlr.de, [†] f.mueller@mx.uni-saarland.de

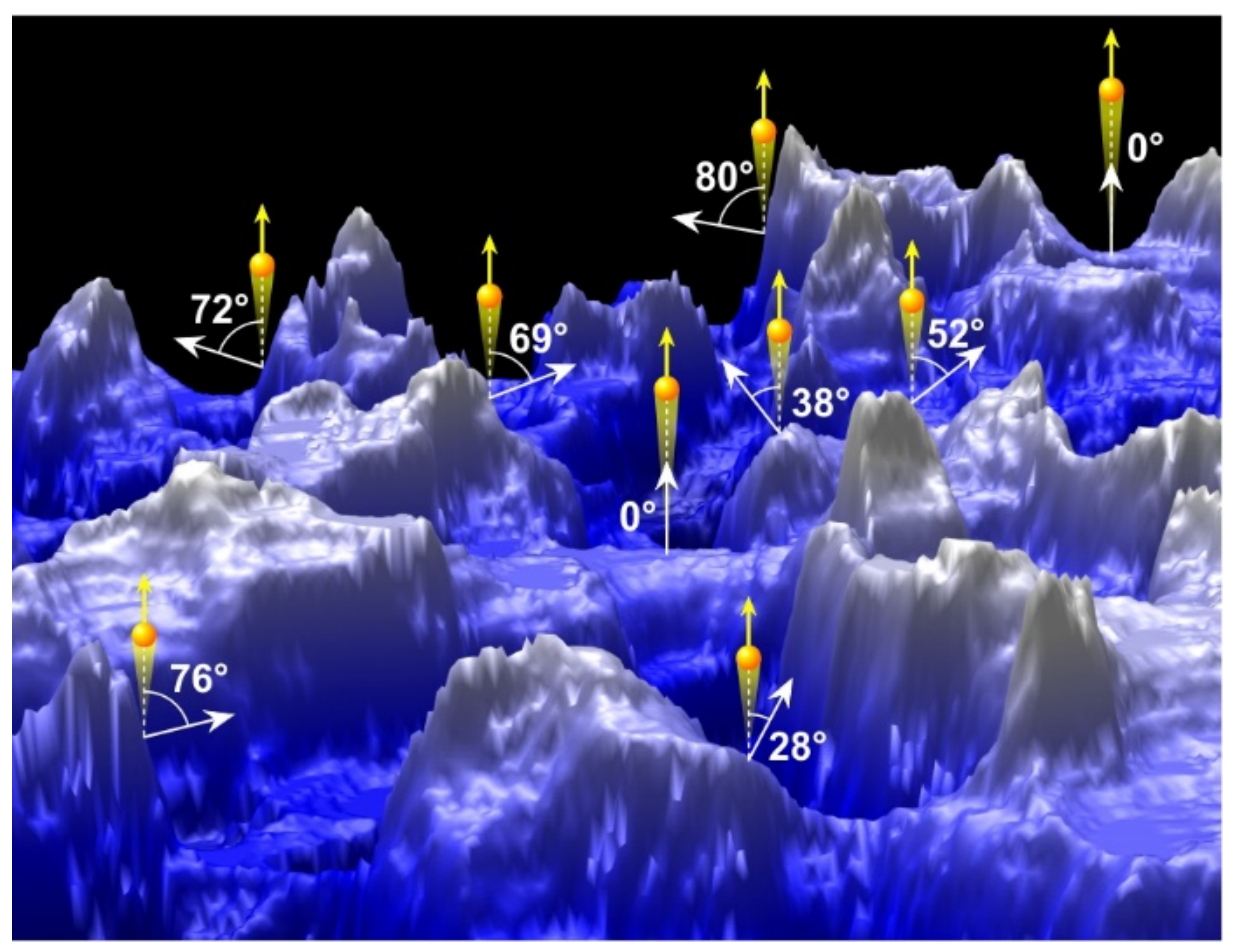

## Abstract

For smooth surfaces, chemical composition can be readily analyzed using various spectroscopic techniques, a prominent example is X-ray photoelectron spectroscopy (XPS), where the relative proportions of the elements are mainly determined by the intensity ratio of the element-specific photoelectrons. However, this analysis becomes more complex for nanorough surfaces like black silicon (b-Si) due to the geometry's steep slopes, which mimic local variations in emission angles. In this study, we explicitly quantify this effect through an integral geometric analysis using Minkowski tensors, correlating XPS chemical data with topographical information from Atomic Force Microscopy (AFM). This approach yields reliable estimates of layer thicknesses for nanorough surfaces. For b-Si, we found that the oxide layer is approximately 50% thicker than the native oxide layer on a standard Si wafer. This study underscores the significant impact of nanoscale geometries on chemical property analysis.

## I. Introduction

In surface science and in materials research, X-Ray Photoelectron Spectroscopy (XPS) is nowadays a standard technique that is mainly applied to analyze the chemical (i.e., elemental) composition of a sample. Compared to other element-sensitive techniques such as Energy Dispersive X-Ray Spectroscopy (EDX), XPS offers the unique advantage of distinguishing different bonding states of atoms through the so-called chemical shifts. In addition, XPS excels by its extreme surface sensitivity since it probes only a few nanometers of the subsurface range of a sample. This surface sensitivity can be further enhanced by increasing the polar angle $\vartheta$ between the surface normal and the entrance axis of the analyzer optics, resulting in an average probing range $\langle z \rangle$ of

$$\langle z \rangle = \lambda \cdot cos\vartheta \qquad (1)$$

where $z$ represents the coordinate perpendicular to the surface and $\lambda$ describes the inelastic mean free path of the particular photoelectron to be probed. Depending on the kinetic energy of the photoelectron, $\lambda$ is typically in the range of a few nanometers [1].

For flat surfaces, the quantitative analysis of XPS data to determine the elemental composition of a sample is straightforward. In the *Appendix* it is demonstrated how the thickness of the native oxide layer of a Si wafer can be determined from angular dependent XPS data using textbook standards [2,3]. However, for (nano)rough surfaces, the advantage of XPS – namely its angular dependent surface sensitivity – seemingly turns into a disadvantage. On a (perfectly) flat surface, the macroscopic surface normal surface normal aligns with the microscopic surface normal of each local area. In contrast on (nano)rough surfaces, the surface normal of local areas can be tilted from the macroscopic surface normal by any angle between 0° and 90° (excluding overhanging features, which are rare and typically not significant for nanorough surfaces). As a result, XPS data collected from (nano)rough samples in normal emission mode (i.e., macroscopic surface normal parallel to the entrance axis of the analyzer optics) represent a superposition of XPS data, locally taken for a broad range of polar angles. Consequently, surface roughness can significantly distort the quantitative analysis of XPS data, particularly in cases where the elemental composition of the surface and subsurface regions is not homogeneous, such as in samples covered by thin layers [4-6].

In this study, we employ a geometric analysis based on Minkowski tensors from integral geometry to analyze the surface topography as probed by AFM. Minkowski tensors (also known as tensor valuations) provide a comprehensive and robust characterization of random geometric structures [7,8]. Their scalar-valued counterparts, the Minkowski functionals, have already been successfully applied to a variety of random heterostructures [8-15]. They were specifically used to quantitatively link surface topography to bacterial adhesion on nano-rough surfaces [16] and motivated a corresponding geometric model [17]. Here, addressing a quite different physical problem, we show how the Minkowski tensors serve as a complementary tool for the evaluation of XPS data from (nano)rough surfaces. Previous studies by Olejnik *et al.* [18] and Zemek et al. [19] have successfully combined XPS and AFM data in proof-of-concept demonstrations, focusing on silicon surfaces with well-known oxide layer thicknesses. However, these studies primarily addressed scenarios with predefined thicknesses, limiting their applicability to cases where the thickness is already known. In contrast, our study advances this approach by aiming to determine an unknown oxide layer thickness on nanorough black silicon (b-Si). By using an angular series of Si-2p XPS data from a smooth Si wafer with a native oxide layer as a reference [16,20], we evaluate how the roughness of black silicon (b-Si), as probed by AFM, influences the distribution of bulk and surface-related spectral features—specifically, the intensity distribution of the bulk-related $Si^0$-2p peak and the surface-related $Si^{4+}$-2p peak of the oxide layer. This approach allows us to estimate the mean thickness of the rough oxide layer with exceptional accuracy, down to the (sub-) Ångström range.

## II. Experimental

### A. Materials

For the angular resolved XPS reference data from a flat surface, a Si wafer in (001) orientation (Siltronic, Burghausen, Germany) was cut into small pieces (approx. 10 mm ´ 10 mm ´ 0.5 mm) which were cleaned with a $CO_2$ snow jet. To meet the constraint of probing the same surface for each polar angle the samples were covered by a tantalum mask of 70 µm thickness and with a slit width of 2 mm (for details, see *Appendix*). The b-Si sample was prepared about 8 years ago according to the synthesis protocol described in ref. [16] and then stored under ambient conditions.

### B. Methods

The XPS experiments were performed with a *Vacuum Generators* ESCALab MkII spectrometer equipped with a hemispherical 180° type EA 15 analyzer by *PreVac*. For excitation the Al-$K_{\alpha}$ radiation (photon energy 1486.6 eV) of an Al/Mg twin anode was used. For survey spectra and for detail spectra of Si-2p the pass energy was set 50 eV and 20 eV, respectively. The pressure was in the range of $5\cdot10^{-10}$ mbar. For quantitative analysis of the intensity distribution of the Si-2p spectra a Shirley background was applied [21].

The surface topography of b-Si was measured in Peak Force Mapping mode with an Icon FastscanBio AFM (Bruker-Nano, Santa Barbara, USA) at ambient conditions using high aspect ratio tips (PFDT750, Bruker-Nano) with a nominal spring constant of 0.4 N/m and load force of 800 pN. The images were taken at five different spots with a 1 µm × 1 µm scan window and a resolution of 512 × 512 pixels. Achieving such a high resolution with pixel sizes just below 2 nm is beneficial for accurately capturing surface roughness. However, in this study, particularly with regard to the XPS data (discussed below), steeper surface slopes contribute less to the overall electron emission. As a result, increasing the resolution even further (e.g., for sub-nanoscale roughness) would not necessarily improve the accuracy of the XPS data. Height channels of the two different fast scan directions (trace/retrace) were treated as separate images. A possible tilt of the surface in the images was removed by subtracting a best fitting plane (‘plain fit, 1st order’). Additionally, a surface reconstruction was performed for each image with different noise suppression thresholds for the estimated tip geometry, following the guidelines from E. Falter *et al.* [22] and J. S. Villarrubia *et al.* [23].

For the Minkowski analysis of the AFM images and their reconstructions, each image was turned into a triangulated surface (with an average area of approx. 5 nm$^2$ per triangle). The latter was constructed via a Delaunay triangulation [24] of the measured points in the $(x, y)$-plane and a subsequent lifting of this planar triangulation to the $z$-values of the AFM data. The Minkowski tensors of the triangulation were then straightforwardly computed using the algorithms detailed in ref. [13]; more specifically, two tensors were evaluated, namely a scalar Minkowski functional $W_1 \coloneqq \int dS$ , i.e., the surface area (also denoted by $S$), and a Minkowski vector $W_1^{01} \coloneqq \int n \, dS$ , i.e., an integral over the surface normal $n$. Additionally, for each triangle at position $(x, y)$ the surface normal and the corresponding angle of inclination $\vartheta(x, y)$ (as defined as the angle between the local surface normal and $z$-axis along the macroscopic surface normal) were determined separately to obtain inclination histograms of the angular distribution of relative surface area, which corresponds to a local version of the Minkowski tensors [25-27].

## III. Results and Discussion

In this study we used a b-Si sample as an exemplary nano-rough surface. Due to its "alpine" roughness with very steep inclinations b-Si excels nowadays as a high-performance material in, e.g., photovoltaic cells and photodetectors [28] and exhibits even a bactericidal effect caused by the spiky structures [16,29].

For b-Si, the quantitative determination of oxide layer thickness via XPS is significantly more complex than for smooth Si wafers (see *Appendix*). Due to the large roughness of b-Si, XPS data collected at a specific polar angle $\vartheta$ (relative to the macroscopic surface normal) always represent angular integrated data.

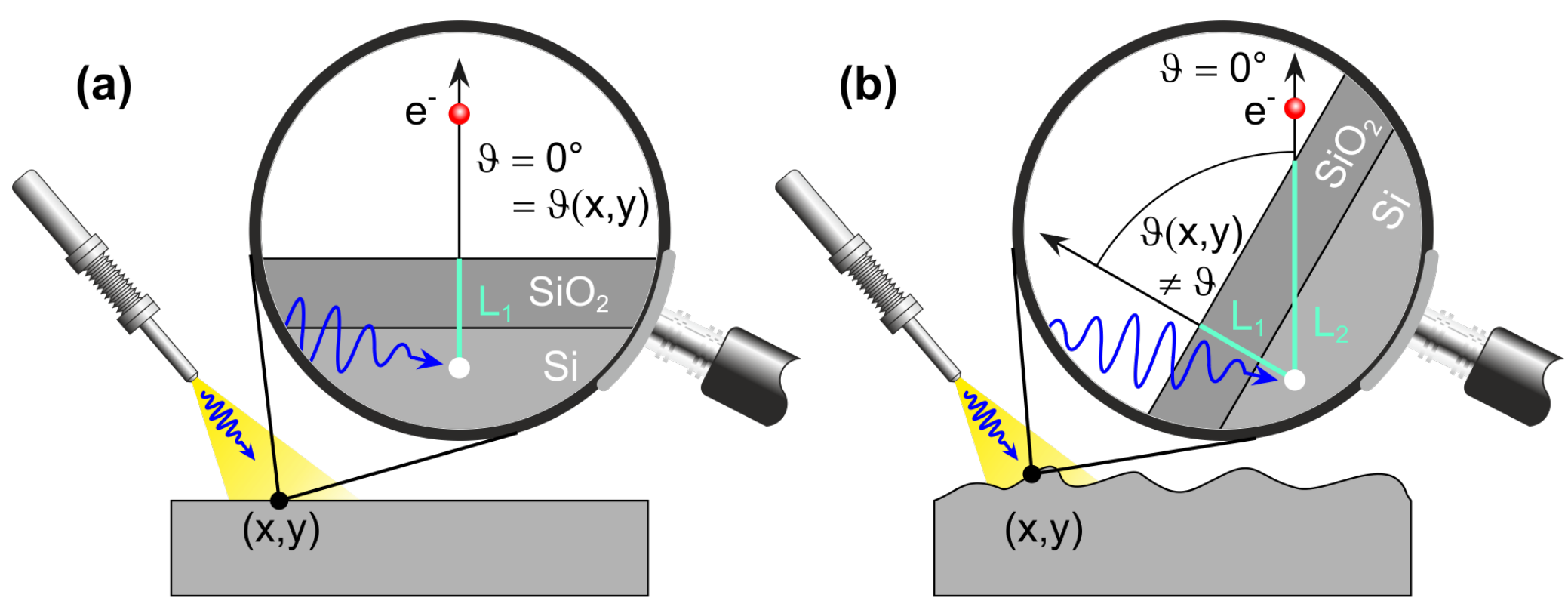

**Figure 1: (a)** Photoemission from Si in the bulk on a smooth Si wafer in normal detection. The local surface normal corresponds to the macroscopic surface normal. **(b)** For a rough surface the local surface normal is tilted by a local angle $\vartheta(x, y)$. The surface position $(x, y)$ contributes more surface sensitively to the normal emission spectrum. For details, see text.

The situation is sketched in Fig. 1. In a normal emission experiment ($\vartheta$=0°) on a smooth Si wafer (Fig. 1a), a Si-2p electron that is emitted in the depth $L_1$ below the surface at any position $(x, y)$ has to pass distance $L_1$ before it crosses the surface into the vacuum. In the same normal emission experiment on b-Si (Fig. 1b), the path length for emission depends on the local inclination at position $(x, y)$ of the surface. A Si-2p electron that is emitted in the same depth $L_1$ below the local surface has to pass the larger distance $L_2 = L_1/\cos\vartheta(x, y)$ (with $\vartheta(x, y)$ being the angle between the local surface normal and the macroscopic surface normal). Although both Si-2p electrons are emitted at the same positions relative to the (local) surface, the Si-2p electron sketched in Fig. 1a contributes with larger probability to the Si-2p peak in XPS than the one in Fig. 1b, as the latter has a longer pathway, increasing the probability of inelastic interactions. In other words: for rough surfaces, the spectral weight shifts towards surface-related features even in a normal emission experiment. For the model system '$SiO_2$-covered Si' this implies that, even with the same oxide layer thickness, the intensity ratio $I(Si^{4c}): I(Si^{0+}\text{-}2p)$ on a rough surface is larger than that probed on a flat surface.

Figure 2a shows a series of Si-2p data for different polar angles $\vartheta$ as taken on a commercial (i.e., flat) Si wafer. The peak at lower binding energy (99.49 ± 0.07 eV) is assigned to the photoemission from the Si-2p orbital of … - $Si^0$- $Si^0$- $Si^0$- … bonded $Si^0$ in the bulk while the peak at higher binding energy (103.42 ± 0.02 eV) is assigned to … - $O^{2-}$- $Si^{4+}$- $O^{2-}$- … bonded $Si^{4+}$ in the oxide layer. With increasing polar angle $\vartheta$ the probing depth decreases, which means that the surface sensitivity increases, and the spectral weight is shifted to the $Si^{4+}$ contribution from the oxide layer.

The intensities $I(Si^{4+})$ and $I(Si^0)$ were determined by fitting two Gaussians to the spectra in Fig. 2a after subtracting a Shirley background [21]. The angular dependency of these intensities as well as the dependency of the rescaled intensities (to eliminate the impact of the used Ta mask of finite thickness, see *Appendix*) are plotted in Fig. 2b. The rescaled intensities are fitted with Eqs. (A6) and (A7) from the *Appendix*, resulting in a thickness of the native oxide layer of 1.27 nm, a value that meets the 1.5 nm of a metallic blue Si wafer [30].

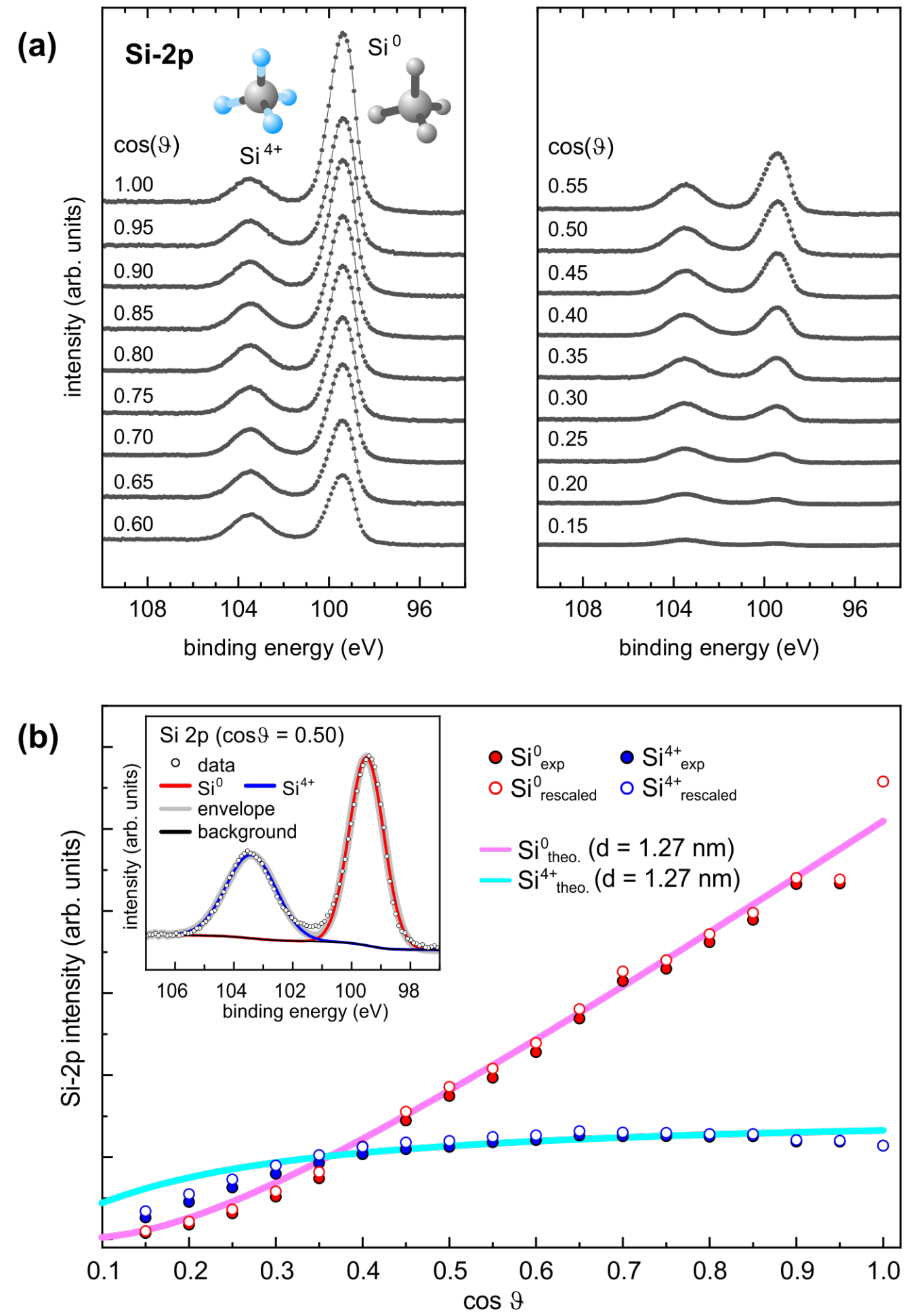

**Figure 2: (a)** Si-2p XPS data from a Si wafer taken at polar angles in equidistant steps $\Delta\cos\vartheta = 0.05$. **(b)** Angular dependence of the $Si^{4+}$ (blue) and $Si^{0}$ (red) intensities (as obtained by fitting the data with two Gaussians on a Shirley background, see example in the inset for $\cos\vartheta = 0.50$). Full dots refer to the experimental data, hollow dots represent rescaled intensities (for eliminating the impact of the mask thickness). The light red and light blue lines represent the fits of the rescaled data according to Eqs. (A6) and (A7) in the *Appendix*, predicting an oxide layer of 1.27 nm thickness.

Figure 3 compares the Si-2p spectrum from a flat Si wafer (Fig. 3a) and a b-Si (Fig. 3b), both taken in normal emission, i.e., $\vartheta=0°$ ($\cos\vartheta=1.00$). In the b-Si spectrum, spectral weight significantly shifts from the $Si^{0}$-2p peak to the $Si^{4+}$-2p peak. The $Si^{4+}$ : $Si^{0}$ intensity ratio in the b-Si spectrum is most comparable to the spectrum in Fig. 3c from the flat Si wafer taken at $\vartheta = 69.5°$ ($\cos\vartheta=0.35$).

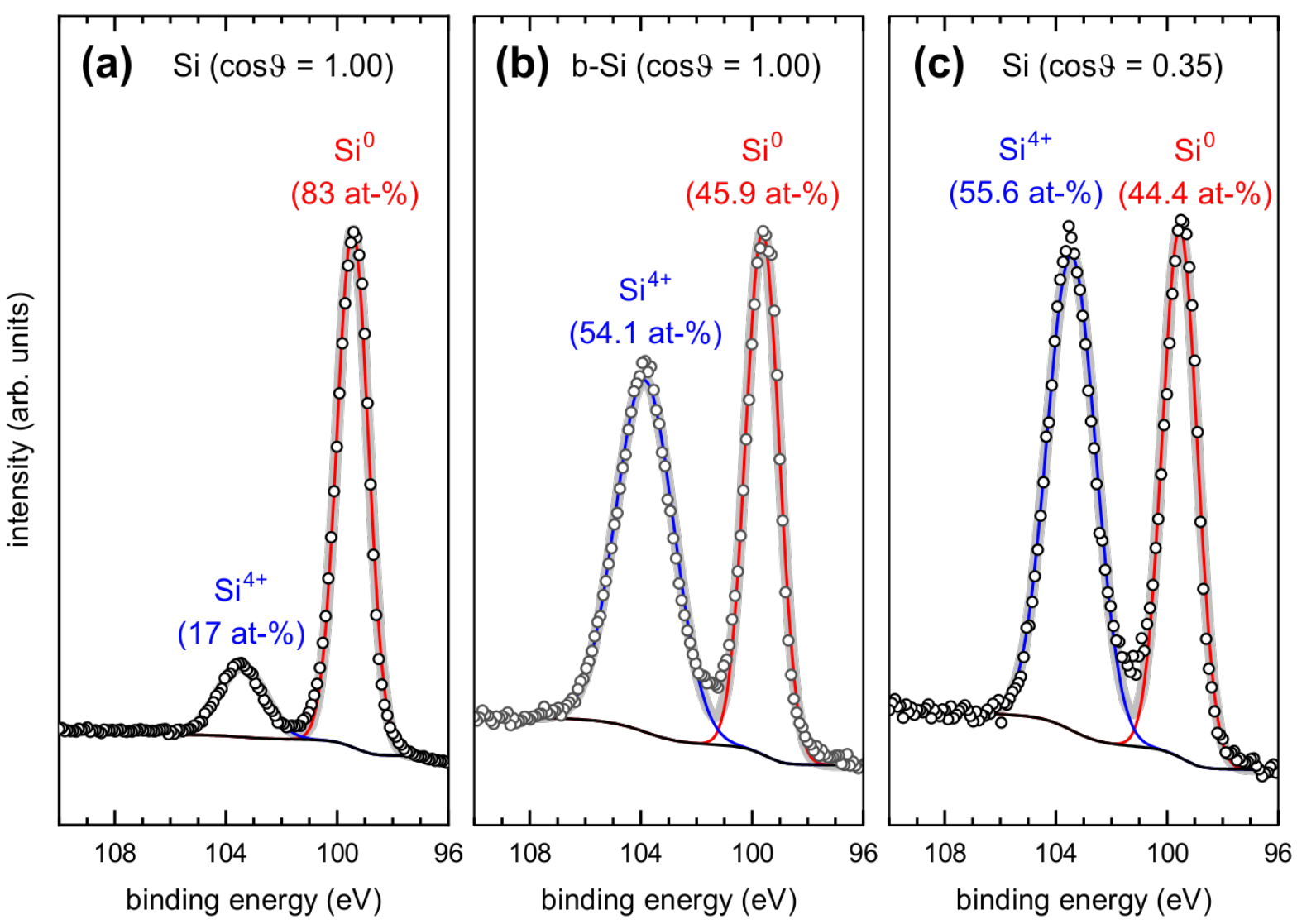

**Figure 3:** Si-2p XPS data from **(a)** a Si wafer at an emission angle ϑ=0° (cosϑ=1.00), **(b)** b-Si at an emission angle ϑ=0° (cosϑ=1.00), **(c)** a Si wafer at an emission angle ϑ=69.5° (cosϑ=0.35). Note that the width of the $Si^{4+}$ peak of b-Si in b) is 25% larger than that of the $Si^{4+}$ peak in c) resulting in different peak heights despite a nearly identical intensity ratio.

At this stage, XPS alone cannot definitively determine whether the increased $Si^{4+}$ : $Si^{0}$ intensity ratio for b-Si is due to an actual increase in oxide layer thickness resulting from the synthesis of this nanorough surface, a mean surface inclination around 70° with an oxide layer comparable to that of a Si wafer, or the interplay of both factors.

To elucidate this question, AFM was used as a complementary technique to probe the topography of the b-Si surface, followed by a geometric analysis based on Minkowski tensors. The AFM images were taken at five different 1 µm × 1 µm scan areas in trace and retrace directions and surface reconstructions were performed for each image by following the guidelines from E. Falter *et al.* [22] and J. S. Villarrubia *et al.* [23]. Figure 4 compares an AFM image and its reconstruction.

The triangulation of the AFM images in Fig. 4a and 4b provides frameworks of triangular tiles in Fig. 4c with each tile at a specific position $(x, y)$ of the surface being tilted by the angle $\vartheta(x, y)$. To describe the local inclination of the surface the $z$-component $n_z$ of the normal vector can be used since it is equal to $\cos\vartheta(x, y)$ for a normal vector $\vec{n}$ of unit length (Fig. 4c). The areas $A(x_i, y_i)$, $A(x_j, y_j)$, … of all tiles with the same inclination, i.e., $\cos\vartheta(x_i, y_i) = \cos\vartheta(x_j, y_j) = \ldots = \cos\vartheta$, add to the total area $A(\cos\vartheta)$, resulting in a histogram $\alpha(\cos\vartheta)$, describing the fraction of surface with inclination $\cos\vartheta$, i.e.,

$$\alpha(cos\vartheta) = \frac{A(cos\vartheta)}{W_1}, \qquad (2)$$

with the Minkowski functional $W_1$ representing the total surface area and $A(cos\vartheta)$ corresponds to the $z$-component of the local Minkowski measure $W_1^{01}$ [25-27].

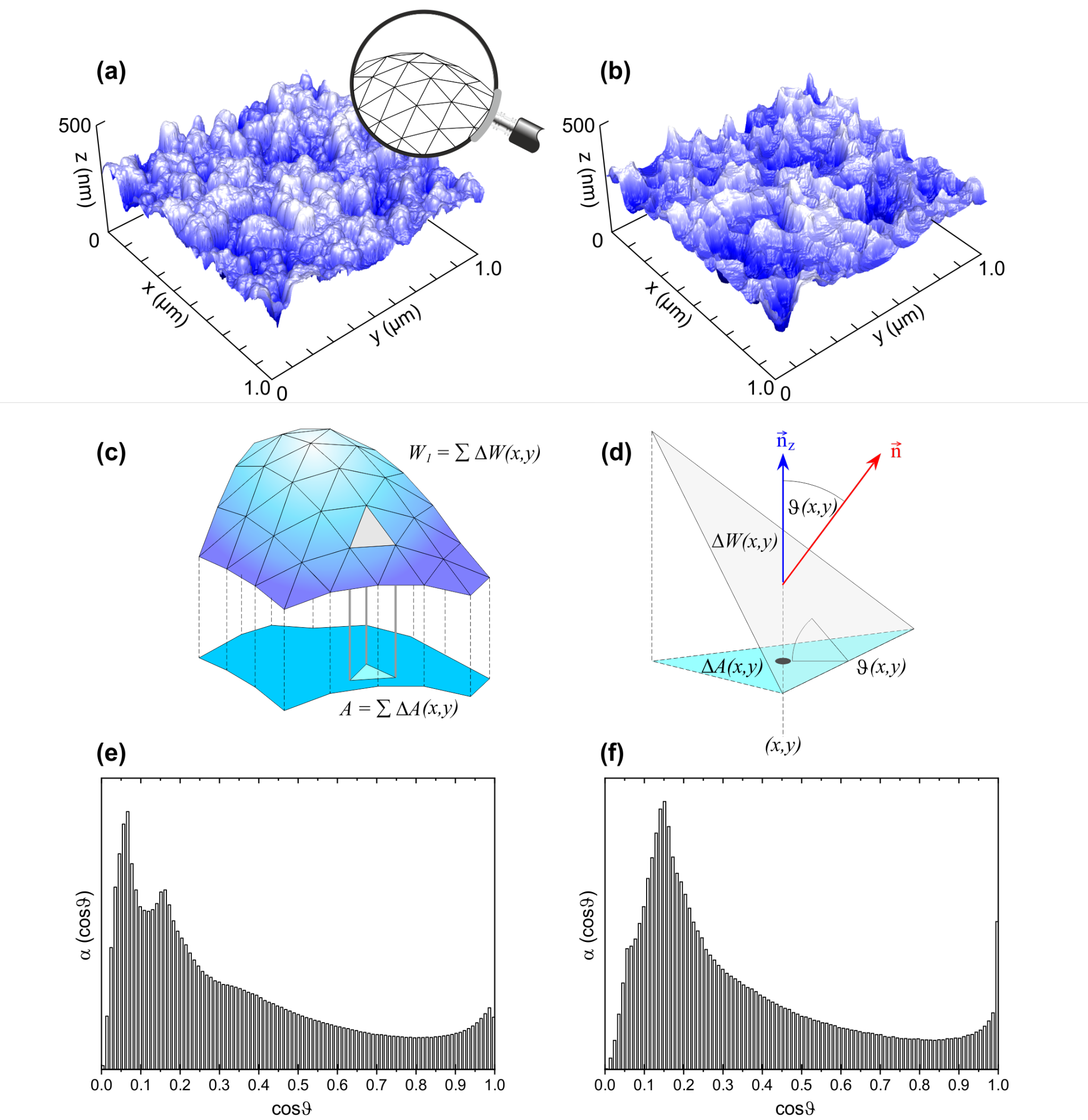

**Figure 4: (a)** 3D representation in real aspect ratio of AFM data of b-Si with 1 µm × 1 µm scan area and a resolution of 512 pixels × 512 pixels. **(b)** Reconstruction of the surface (for details, see text). **(c)** Scheme of a 3D overall surface (Minkowski measure $W_1$) as a superposition of locally flat surfaces. **(d)** The z component of the normal vector (of unity length) of a tilted surface is a direct measure for the inclination of the surface, $n_z(x,y) = \cos\vartheta(x,y)$. **(e)** Angular distribution of relative surface areas as extracted from the original AFM image in (a) and **(f)** as extracted from the reconstructed AFM image in b). In both histograms all values add to unity.

Figures 4e and 4f show the histograms of $\alpha(\cos\vartheta)$ for the original measured and reconstructed AFM images, respectively. There are two salient features. First, an additional maximum at a comparable high slope (at $\cos 86.6° = 0.06$) in the histogram from the original AFM images in Fig. 4e. The most likely explanation for this is caused by noise and especially image errors (i.e., an overshooting in the height signal). This problem of the original data can be avoided by the reconstruction of the surface (the peak disappears in the histogram of the reconstructed AFM image in Fig. 4f). Simultaneously, the position of the second maximum remains at around $0.16 \approx \cos 81°$, which gives confidence in the reconstruction. Secondly, the histogram for the reconstructed image in Fig. 4f exhibits a sharp peak at $\cos 0° = 1$ (i.e., for exactly horizontal triangles), which can be explained by a too-blunt shape of the estimated tip that eradicates structures smaller than the tip size (resulting in flat triangles). Note that noise and image errors strongly affect the tip estimation and unavoidably limit the accuracy of the reconstruction [22,23]. The algorithm for tip reconstruction and subsequent surface deconvolution uses therefore a threshold value for disregarding noise as an input value from the user. Choosing too high values for this noise threshold leads, however, also to reconstruction artifacts which appear as flattened areas in the image. Falter et al. [22] report about a method to determine a best choice for this threshold. With the resulting threshold, the mentioned flattened areas are can also be mostly avoided. Applying their method to the b-Si surfaces leads, however, to an ambiguous result, i.e., several of these "best choices" were obtained for each image. As it is not possible at this point to determine a "correct" tip reconstruction and in order to avoid personal bias, the histograms for $\alpha(\cos\vartheta)$ were compiled from four to five deconvoluted surfaces per image (as weighted average). Moreover, flattened areas (with $\vartheta$ = 0) were removed from the histogram as they are clearly reconstruction artifacts at spots where high aspect ratio peaks are expected. Although these areas were only ca. 3% of the overall surface, they'd lead to an overestimation of the oxide layer thickness as they virtually contribute with a high $Si^0$ signal instead of a rather low one, which would be expected from sharp needle-like surface structures.

With the AFM-derived distribution of local surface inclinations in Figs. 4e and 4f and with the theoretically calculated angular distributions of the $Si^0$-2p and $Si^{4+}$-2p intensities (as shown in Fig. 2b for d = 1.27 nm), it is straightforward to estimate the thickness of the oxide layer of b-Si. Fig.5 depicts graphically the calculation: In Fig. 5a and 5d, the emission angle dependency of the $Si^0$-2p and $Si^{4+}$-2p intensities are plotted for a certain thickness $d$ of the oxide layer of a flat Si wafer according to Eqs. (A6) and (A7). To take the roughness of b-Si into account, these angular functions are weighted with the angular distribution of the tilted areas from Fig. 4e and 4f. Here, the distribution as derived from the original AFM data in Fig. 4e are displayed as an example in Figs. 5b. The overall areas of the inclination-weighted $Si^0$-2p and $Si^{4+}$-2p intensities in Fig. 5c and 5e are then direct measures for the $Si^0$-2p and $Si^{4+}$-2p intensities of b-Si:

$$I(bSi^0 2p) \sim \sum_{cos\vartheta} \{\alpha(cos\vartheta) \cdot eq.(A6)\} \qquad (3)$$

$$I(bSi^{4+} 2p) \sim \sum_{cos\vartheta} \{\alpha(cos\vartheta) \cdot eq.(A7)\} \qquad (4)$$

Varying the film thickness d and thus the functions shown in Figs. 5a and 5d, a film thickness dependency of the $Si^0$-2p intensity ratio is received and shown in Fig. 6. Here, the $Si^0$-2p ratio is calculated as

$$R^0_{calc.} = \frac{I(bSi^0 2p)}{I(bSi^0 2p) + I(bSi^{4+} 2p)} \qquad (5)$$

From this plot, the film thickness of the b-Si can directly be read by using the experimentally determined $Si^0$-2p intensity ratio of

$$R^{0}_{exp.} = \ 45.9 \ \pm 0.7 \text{ at-\%}.$$

Computationally, this was realized by an optimization procedure with varying d as schematically sketched in Fig. 5 in terms of a computational loop.

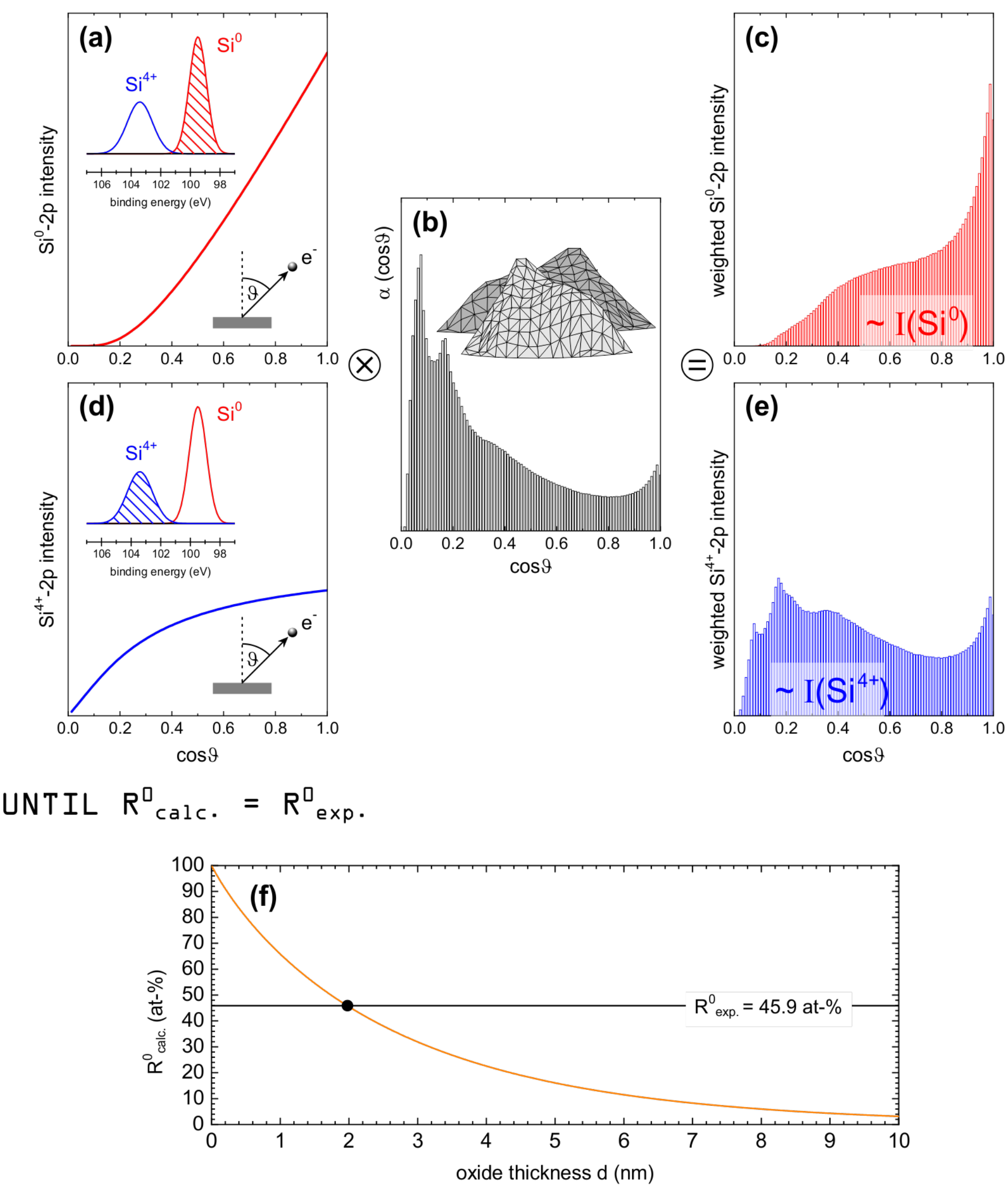

**Figure 5: (a)** Angular distribution of $Si^0$-2p intensity according to eq. (A6) for a specific oxide thickness d. **(b)** Angular distribution of inclination (here from original AFM data in Fig. 4e). **(c)** Angular distribution of inclination weighted $Si^0$-2p intensity. **(d)** Angular distribution of $Si^{4+}$-2p intensity according to eq. (A7) for same oxide thickness d as in (a). **(e)** Angular distribution of inclination weighted $Si^{4+}$-2p intensity. **(f)** The parameter d in (a) and (d) is varied until the areas of the histograms in (c) and (e) meet the condition $R^0_{calc.} = R^0_{exp.}$.

In summary: Figs. 5a and 5d is input based on angular resolved XPS on flat Si wafers. Fig. 5b is input based on the mapping of topography of rough b-Si by AFM. Figs. 5c and 5e is output for angular resolved XPS on rough b-Si, i.e., an experiment that cannot be performed in reality.

For a sample covered with an adsorbate of thickness $d_{ads.}$ Eqs. (A13) and (A14) have to be used in the same way, i.e.,

$$I_{ads.}(bSi^{0}2p) \sim \sum_{cos\vartheta} \{\alpha(cos\vartheta) \cdot eq.(A13)\} \qquad (6)$$

$$I_{ads.}(bSi^{4+}2p) \sim \sum_{cos\vartheta} \{\alpha(cos\vartheta) \cdot eq.(A14)\} \qquad (7)$$

Figure 6 compares the calculated dependence of the $Si^{0}$-2p ratio from thicknesses of the silicon oxide layer using different assumptions for the sample: (i) a flat Si wafer (purple line), i.e., Eqs. (A6) and (A7) are used without weighting, (ii) b-Si using the original AFM data for weighting (orange) and (iii) b-Si using the deconvoluted surface data after tip reconstruction (green). Additionally, for these three weighting methods, an adsorbate was assumed using Eqs. (A13) and (A14), yielding the dashed lines in Fig. 6, respectively. The adsorbate was thereby simulated by a 5 Å layer of graphite (using the electron mean free paths of Eqs. (A15) and (A16)). Moreover, the experimentally determined $Si^{0}$-2p ratio is shown as black horizontal line.

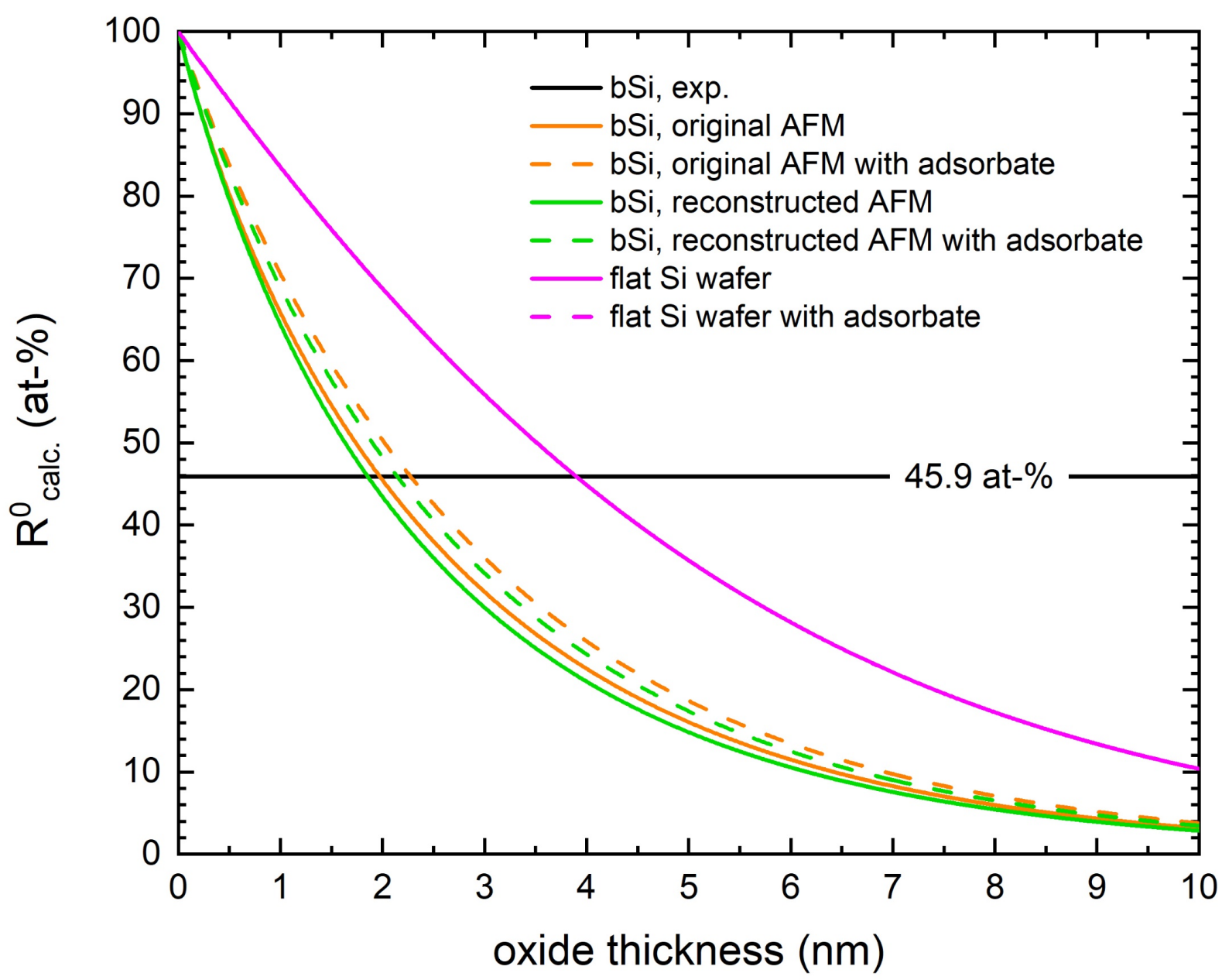

**Figure 6:** Calculated $Si^{0}$-2p intensity ratio in dependence of oxide thickness of b-Si for clean and adsorbate (0.5 nm graphite) surfaces using the original AFM data and the reconstructed AFM data. In addition, the thickness dependence of a flat Si wafer is shown. The values of thickness are listed in table I.

The obtained values for the thicknesses are also listed in table I including the uncertainty range. It is interesting to note that for a flat Si wafer the adsorbate has nearly no influence on the results, i.e., the dashed and the solid line coincide. This is due to the additional attenuation factors in Eq. (A13) and Eq. (A14) being nearly the same.

For b-Si, the surface angle distributions in Fig.4e-f are very different, yet their impact on the oxide thickness is rather marginal, i.e., full lines as well as dashed lines in Fig. 6 nearly coincide, resulting in differences of oxide thickness ≤ 5 %. Contrasting the obtained values from flat and b-Si, however, shows that a false assumption for the surface topography may lead to a drastic overestimation of the oxide film thickness.

| | | $d \pm \Delta d$ (nm) |
|---|---|---|
| clean | original AFM | $1.98 \pm 0.04$ |
| | reconstructed AFM | $1.89 \pm 0.04$ |
| | flat Si wafer | $3.90 \pm 0.07$ |
| with 5 Å adsorbate of graphite | original AFM | $2.26 \pm 0.04$ |
| | reconstructed AFM | $2.15 \pm 0.04$ |
| | flat Si wafer | $3.90 \pm 0.07$ |

**Table I:** Oxide thickness for b-Si with a $Si^0$ ratio of 45.9 at-% when using the angular distributions from the original and the reconstructed AFM with and without adsorbate (graphite, 0.5 nm). The errors for d are obtained by calculating the respective thickness values for the error interval limits of the $Si^0$-2p intensity ratio as obtained from the peak fitting of the XPS data of b-Si in Fig. 3b.

The method presented above is a prime example of how Minkowski functionals can be used to combine two methods like AFM and XPS to gain access to the (oxide) layer thickness via the calculation of the distribution of orientation/inclination. Moreover, this method can be used whenever intuitive, accessible geometrical measures are needed to support the data evaluation of spectroscopic techniques. Accessible in this context means, that there are different types of geometrical measures which can be calculated via Minkowski functionals. Directly accessible are all additive shape information [7] such as volume, surface area, Euler characteristic etc. Some non-additive shape information is also indirectly accessible, e.g., (bounds on) percolation thresholds [31].

## IV. Conclusion

This study demonstrates that surface topography can significantly influence the characterization of surface-related properties, even when the experimental techniques used are not directly related to surface topography. In the case of XPS, a highly surface-sensitive spectroscopic technique, the evaluation of surface stoichiometry and the near-subsurface range can be substantially distorted by increasing surface roughness. For any polar angle in the experiment, XPS data from rough surfaces inherently represent angularly integrated data. For b-Si, a model surface with extreme roughness, XPS data of Si-2p taken in normal emission suggest an oxide thickness nearly three times larger than that of the native oxide. However, when accounting for the impact of surface roughness, the actual oxide thickness of b-Si is found to be approximately 49 - 56% larger (1.89 nm or 1.98 nm vs. 1.27 nm for clean surfaces) or approximately 69 - 78% larger (2.15 nm or 2.26 nm vs. 1.27 nm for adsorbate-covered surfaces) than that of a reference Si wafer.

Characterizing the surface topography with Minkowski measures (here: AFM-derived surface $W_1$ and its distribution of orientation/inclination) proves to be a valuable tool for identifying roughness-induced changes in spectral features. This Minkowski analysis directly maps the roughness of the sample without requiring simulations using basic geometric models like hemispheres, pyramids, or cones.

The combined XPS/AFM analysis presented in this study enables precise determination of layer thicknesses on such rough samples where other techniques, such as ellipsometry, fail. With standard X-ray sources like Al-$K_\alpha$ or Mg-$K_\alpha$ radiation, this method is limited to layers with thicknesses of only a few nanometers since the electron mean free paths are typically also in the range of a few nanometers. However, samples with layers of larger thickness can be analyzed using HXPS (Hard X-ray photoelectron spectroscopy), which offers increased electron mean free paths but at the cost of significantly reduced cross sections. This disadvantage can be mitigated by using synchrotron radiation to increase photon flux by several orders of. The effectiveness of the combined (H)XPS/AFM method depends on the accurate knowledge of electron mean free paths, which vary with kinetic energy and material properties. When precise data are unavailable, the so-called universal curves by Seah and Dench [1] can be used to estimate these paths. This study not only advances the understanding of surface characterization on nanorough materials but also highlights the potential of integrating geometric and spectroscopic analyses for more accurate and reliable measurements, paving the way for further innovations in surface science and materials research.

## Acknowledgement

This work was supported by the German Research Foundation (Deutsche Forschungsgemeinschaft, DFG) via the Priority Program SPP 2265 “Random Geometric Systems” and via the Collaborative Research Center CRC/SFB 1027 “Physical modeling of non-equilibrium processes in biological systems”. M.A.K. acknowledges funding and support by the DFG through SPP 2265 under Grant Nos. KL 3391/2-2, WI 5527/1-1, and LO 418/25-1, as well as by the Initiative and Networking Fund of the Helmholtz Association through the Project “DataMat.”

## Author contributions

F.M. developed conceptional idea. J.U.N., K.J., M.A.K., F.M. designed research. F.N. produced b-Si samples. J.U.N., F.N. performed AFM measurements. J.U.N., F.M., S.G., T.F. performed XPS measurements. J.U.N, H.H., M.A.K. performed Minkowski analyses. A.W. developed computational code for Minkowski analysis. J.U.N., H.H., M.A.K., F.M. analyzed data. J.U.N., H.H., K.J., M.A.K., F.M. wrote the manuscript with the help of all other authors.

## Appendix: Quantitative analysis of XPS data

This section describes the quantitative analysis of XPS data in terms of calculating the elemental composition of a sample (i.e., stoichiometry) according to the textbook standards [2,3]. In XPS the intensity (= number of photoelectrons) is determined by several factors.

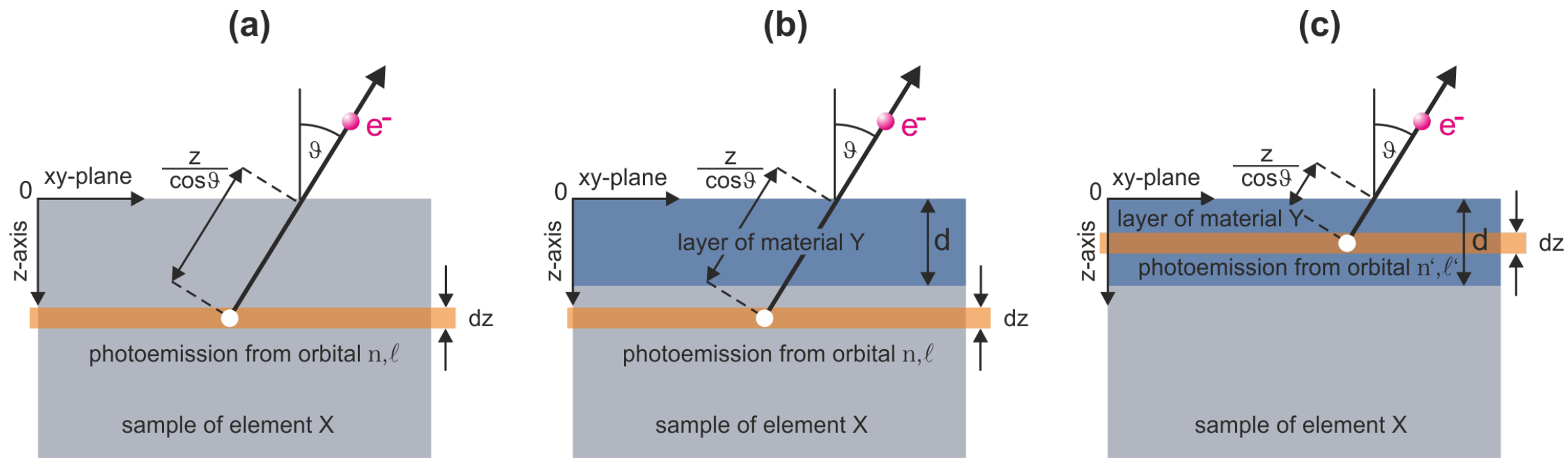

**Figure 7:** Scheme of **(a)** the photoemission process from element X for a uniform sample of element X, **(b)** the photoemission process from element X for a uniform sample of element X that is covered by a layer from material Y of thickness d, **(c)** the photoemission process from an element of layer Y. For details, see text.

For a sample of a homogeneous material (Fig. 7a) photoelectrons that are emitted from the orbital or core level (n, $\ell$) of atoms of element X within a slice of thickness $dz$ in the depth $z$ below the surface contribute to intensity as

$$dI(X,n,\ell) \sim \sigma_{X,n,\ell} \cdot \rho_X \cdot T_{X,n,\ell} \cdot A(\vartheta) \cdot \exp\left(-\frac{z}{\lambda_{X,n,\ell} \cdot cos\vartheta}\right) dz \qquad \text{(A1)}$$

with

$\sigma_{X,n,\ell}$ — photoemission cross section of orbital (n, $\ell$) of element X, depending on photon energy used for excitation. n = principal quantum number, $\ell$ = angular momentum

$\rho_X$ — atomic density of element X

$T_{X,n,\ell}$ — transmission of the analyzer setup depending on kinetic energy of electrons emitted form orbital (n, $\ell$) of element X

$A(\vartheta)$ — detected area of sample with the polar angle $\vartheta$ between the (macroscopic) surface normal of the sample and the lens axis of the analyzer optics

$\lambda_{X,n,\ell}$ — inelastic mean free path in material X depending on kinetic energy of electrons emitted form orbital (n, $\ell$)

The exponential damping factor considers that the larger the depth at which the photoemission takes place the larger the probability that the corresponding photoelectrons contribute to the inelastic background rather than to the primary signal (peak X-n$\ell$, e.g., *Si-2p*) due to inelastic interactions.

In case of a homogeneous Si sample (i.e., Si wafer without oxide layer, Fig. 7a) the total yield of $Si^0$-$2p$ electrons is obtained by integration of Eq. (A1) from $z = 0$ (surface) to $z = \infty$ (since the thickness of a 0.5 mm Si wafer exceeds the inelastic mean free paths (typically 1-2 nm) by several orders of magnitude)

$$I(Si^0 2p) \sim \sigma_{Si^0 2p} \cdot \rho_{Si} \cdot T_{Si^0 2p} \cdot A(\vartheta) \cdot \lambda_{Si^0 2p} \cdot cos\vartheta \qquad \text{(A2)}$$

The inelastic mean free path is further specified as

$$\lambda_{Si^0 2p}(Si^0)$$

meaning the average distance $Si^0$-$2p$ photoelectrons can travel in the material Si until there is an inelastic interaction.

Figure 7b sketches the situation for a Si wafer with an additional layer of native oxide ($SiO_2$) of thickness $d$. The total yield of $Si^0 2p$ photoelectrons is reduced because in the oxide layer there is no source for $Si^0$-$2p$ electrons but there is an additional damping of Eq. (A2) when the $Si^0$-$2p$ photoelectrons inelastically interact in the $SiO_2$ layer, i.e.,

$$I(Si^0 2p) \sim \sigma_{Si^0 2p} \cdot \rho_{Si} \cdot T_{Si^0 2p} \cdot A(\vartheta) \cdot \lambda_{Si^0 2p}(Si^0) \cdot cos\vartheta \cdot exp\left(-\frac{d}{\lambda_{Si^0 2p}(SiO_2) \cdot cos\vartheta}\right) \qquad \text{(A3)}$$

In the damping factor the electron mean free path

$$\lambda_{Si^0 2p}(SiO_2)$$

considers that now $Si^0$-$2p$ photoelectrons are inelastically scattered in a different material.

The total yield of $Si^{4+}$-$2p$ photoelectrons as emitted from the $SiO_2$ layer in Fig. 7c is obtained by integrating Eq. (A1) from 0 to the layer thickness $d$.

$$I(Si^{4+} 2p) \sim \sigma_{Si^{4+} 2p} \cdot \rho_{SiO_2} \cdot T_{Si^{4+} 2p} \cdot A(\vartheta) \cdot \lambda_{Si^{4+} 2p}(SiO_2) \cdot cos\vartheta \cdot \left(1 - exp\left(-\frac{d}{\lambda_{Si^{4+} 2p}(SiO_2) \cdot cos\vartheta}\right)\right) \qquad \text{(A4)}$$

In Eqs. (A3) and (A4) some of the parameters can be combined into a common factor:

(i) For $Si^0$-$2p$ photoelectrons from the bulk and for $Si^{4+}$-$2p$ photoelectrons from the oxide layer the photoemission cross sections $\sigma_{Si^0 2p}$ and $\sigma_{Si^{4+} 2p}$ are the same [32,33].

(ii) In this study the photoemission intensities of $Si^0 2p$ and $Si^{4+} 2p$ are compared and the difference in their kinetic energies of these electrons is less than 0.1 % of their kinetic energies (when excited with Al-$K_\alpha$ radiation). Since the transmission factors are ~ $E_{kin}^{\beta}$ (with $\beta$ being a spectrometer specific constant) the ratio of the transmission factors $T_{Si^0 2p}$ : $T_{Si^{4+} 2p}$ is $1.001^{\beta}$ with is equal 1 (independent on $\beta$). Therefore, both factors can be equated.

(iii) The area $A(\vartheta)$ that contributes the photoemission intensity depends on the size of the sample. In case that the area is larger than the field of view (FOV) of the analyzer setup (as given by the particular choice of iris apertures and slit widths) the area increases with increasing polar angle $\vartheta$ (see Fig. 8a). In case that the area is smaller than the FOV, the area initially remains constant to a certain polar angle (see Fig. 8b).
However, in both cases, at very large angles, there is the risk that besides the actual surface (tilted by the angle $\vartheta$) also the edge area (tilted by the angle 90°-$\vartheta$) also contributes to the photoemission signal. For large polar angles, the latter contribution falsifies Eqs. (A3) - (A4) to a large extent.
Therefore, it is necessary to define the size of the sample´s surface area by using a mask. When using a mask, the surface area that is probed also depends on the polar angle $\vartheta$. In normal emission ($\vartheta$=0°) the full area is probed (Fig. 9a) while for increasing polar angles a part of the area is hidden due to the finite thickness of the mask. From Fig. 9b it is straightforward to rescale the experimental photoemission intensities to the intensities for a zero-thickness mask via

$$I(\vartheta) = I_{exp}(\vartheta) \cdot \frac{1}{1-t/w \cdot tan\vartheta} \qquad \text{(A5)}$$

with t and $w$ representing thickness and slit width of the mask, respectively.

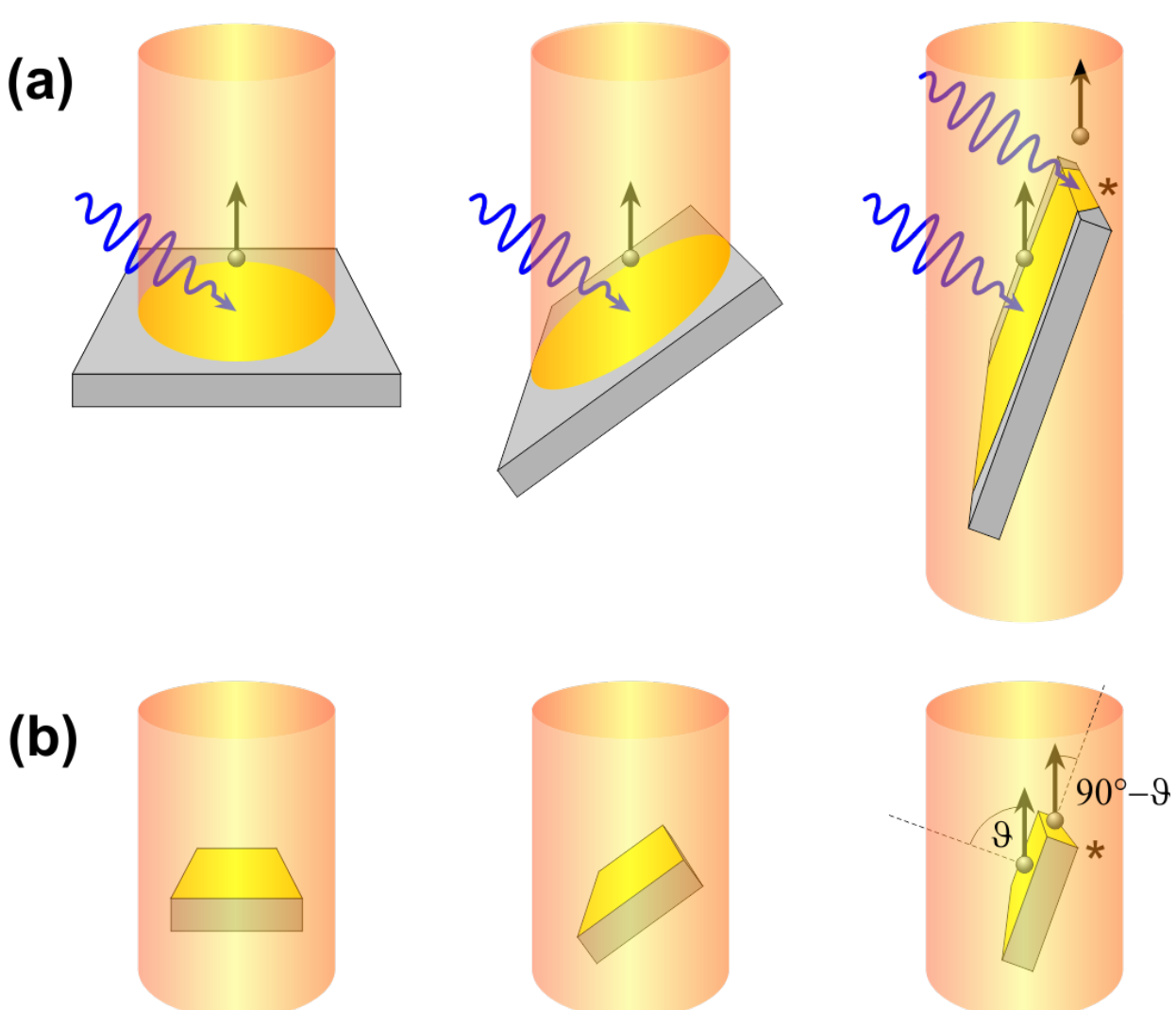

**Figure 8:** Variation of the probed surface area in dependence of the polar angle $\vartheta$ for **(a)** A sample larger than the field of view of the analyzer system (here represented by a cylindrical tube for simplicity) and **(b)** for a sample smaller than the field of view. In both cases also photoemission intensities from the edges (*) can contribute to the overall intensities. For details, see text.

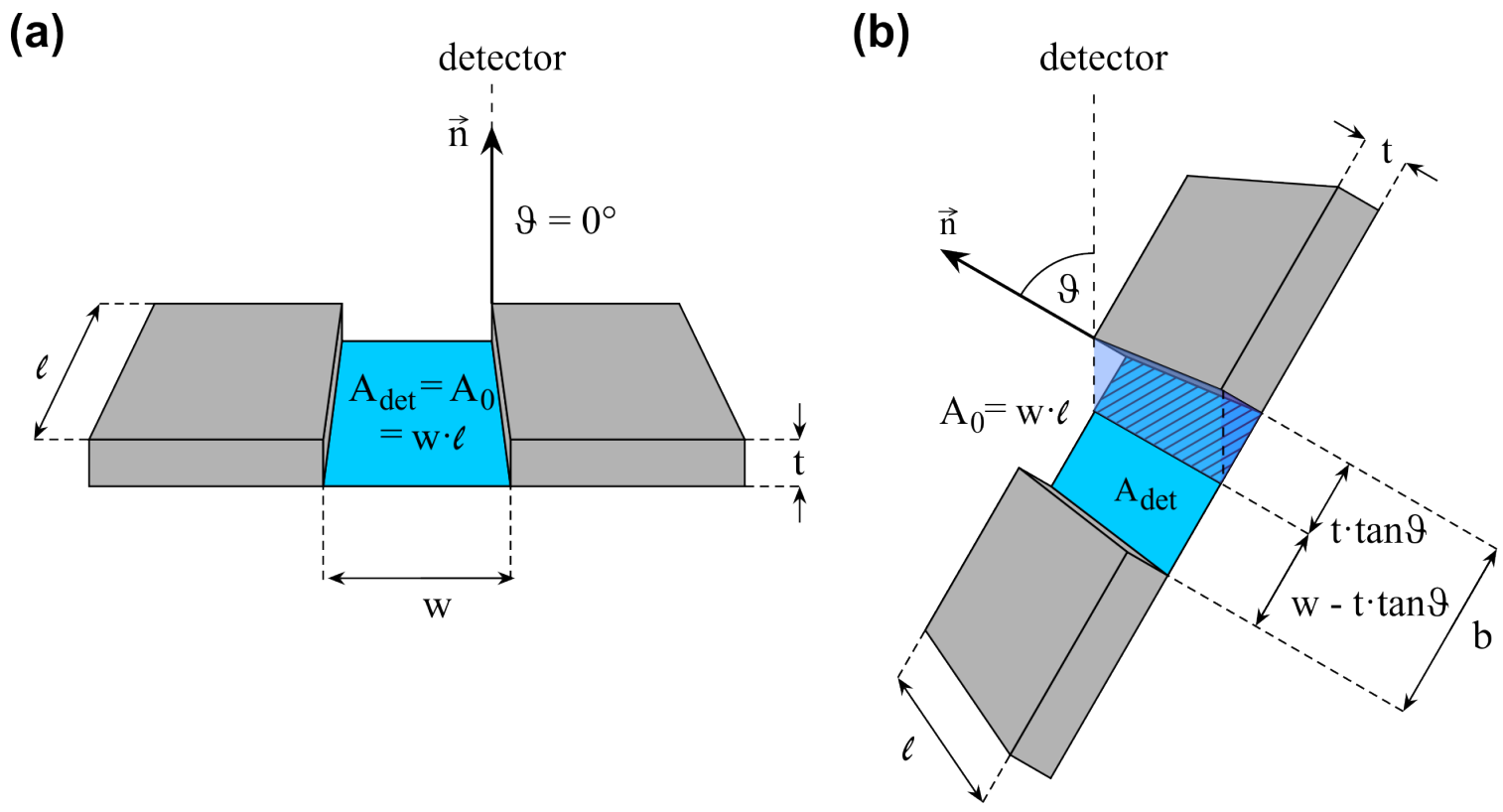

**Figure 9:** Impact of a mask of finite thickness $t$ on the probed surface area. **(a)** In normal emission ($\vartheta$ = 0°) the full area is probed. **(b)** For non-normal emission ($\vartheta$ > 0°) only the fraction $1 - t/w \cdot \tan\vartheta$ of the full area is probed ($t$: thickness of mask, $w$: width of mask). For details, see text.

When designing the experiments according to (i) - (iii), the photoemission cross sections, the transmission functions and the (now $\vartheta$-independent) area can be combined in a common factor $f$ in Eqs. (A3) - (A4) and one gets two equations with only two free parameters to describe the intensity distributions of the $Si^0$-$2p$ and $Si^{4+}$-$2p$ intensities in an angular series.

$$I(Si^0 2p) \;=\; \boldsymbol{f} \cdot \rho_{Si} \cdot \lambda_{Si^0 2p}(Si^0) \cdot cos\vartheta \cdot exp\left(-\frac{\boldsymbol{d}}{\lambda_{Si^0 2p}(SiO_2) \cdot cos\vartheta}\right) \qquad \text{(A6)}$$

$$I(Si^{4+} 2p) \;=\boldsymbol{f} \cdot \rho_{SiO_2} \cdot \lambda_{Si^{4+} 2p}(SiO_2) \cdot cos\vartheta \cdot \left(1 - exp\left(-\frac{\boldsymbol{d}}{\lambda_{Si^{4+} 2p}(SiO_2) \cdot cos\vartheta}\right)\right) \qquad \text{(A7)}$$

The atomic densities $\rho_{Si}$ and $\rho_{SiO_2}$ (in terms of number of Si atoms per volume) can be obtained by the mass densities, i.e., 2.34 g/cm$^3$ for Si and 2.65 g/cm$^3$ for $SiO_2$ [34,45], and the molar masses M(Si) = 28.08 g/mol and M($SiO_2$) = 60.08 g/mol:

$$\rho_{Si} = \frac{2.34\ g/cm^3}{28.08\ g/mol} = 0.0833\ \frac{mol}{cm^3} \qquad \text{(A8)}$$

$$\rho_{SiO_2} = \frac{2.65\ g/cm^3}{60.08\ g/mol} = 0.0441\ \frac{mol}{cm^3} \qquad \text{(A9)}$$

The values of the inelastic mean free paths in Si and in $SiO_2$ are taken from an online calculator [36] that refers to the study by Tanuma *et al.* [37]. For a photon energy of 1486.6 eV (Al-K$_\alpha$) and a spectrometer work function of 4.0 eV the kinetic energies of 1383.1 eV and 1379.4 eV were used for $Si^0$-$2p$ (at binding energy of 99.5 eV) and $Si^{4+}$-$2p$ (at binding energy of 103.2 eV), respectively. Ref. [36] then provides for the electron mean free paths in Eqs. (A6) and (A7)

$$\lambda_{Si^0 2p}(Si^0) = 3.082\ nm \qquad \text{(A10)}$$

$$\lambda_{Si^0 2p}(SiO_2) = 3.745\ nm \qquad \text{(A11)}$$

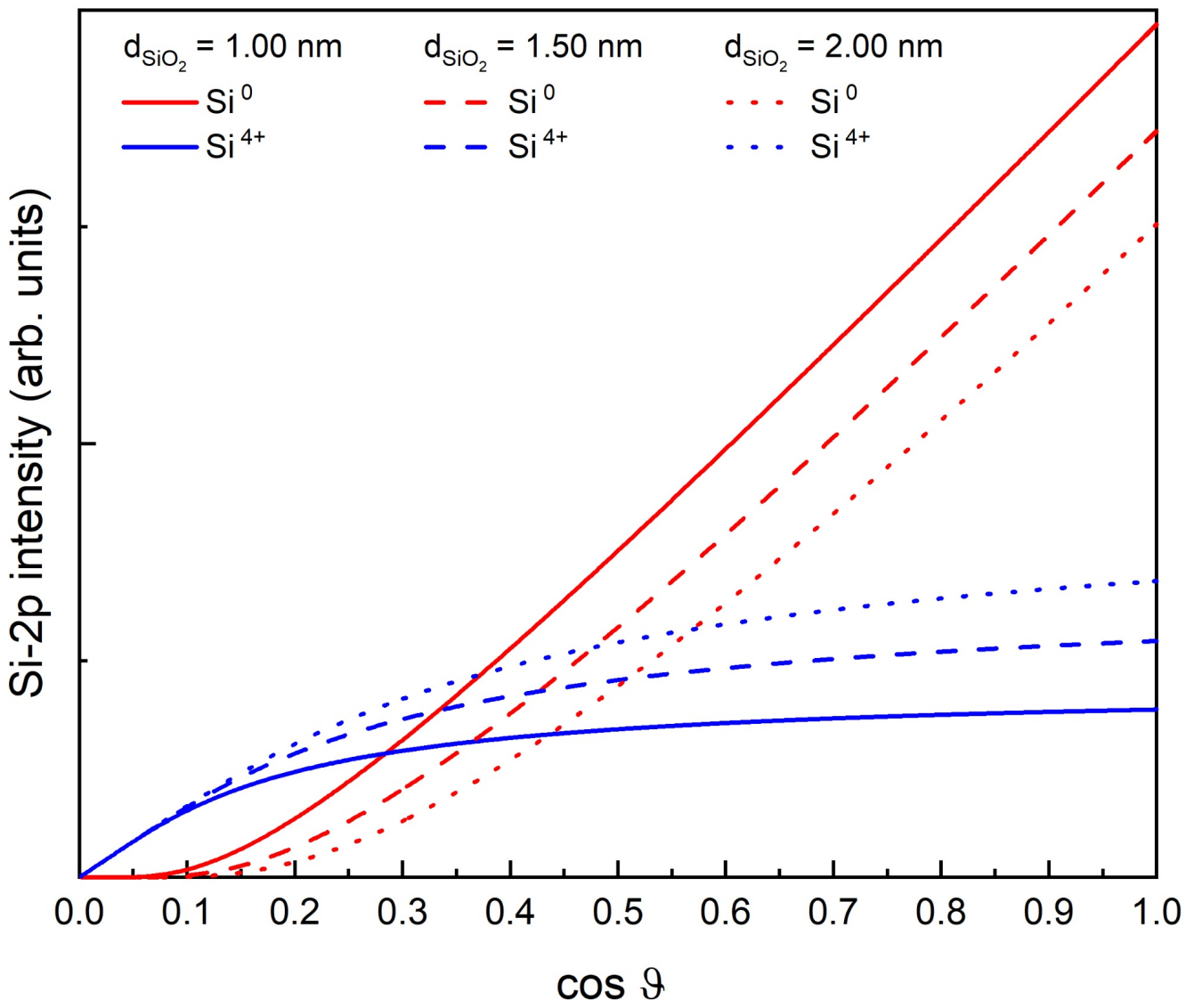

**Figure 10:** Angular distribution of the $Si^0$-$2p$ photoemissions intensities (red) and the $Si^{4+}$-$2p$ (blue) photoemissions intensities from a flat Si wafer with oxide thickness of 1.00 nm (solid lines), 1.50 nm (dashed lines) and 2.00 nm (dotted lines).

$$\lambda_{Si^{4+}2p}(SiO_2) = 3.737\ nm \qquad \text{(A12)}$$

Figure 10 show the distribution of the $Si^0$-$2p$ and the $Si^{4+}$-$2p$ photoemission intensities in dependence of the polar angle according to Eqs. (A6) - (A7) for a flat Si wafer using the parameters (A8) - (A12) for three different thicknesses $d$. Equations (A6) and (A7) are the main input for the estimation of the oxide thickness of rough surfaces. In Fig. 2 it is demonstrated that the model described here reproduces very well the angular intensity distributions as observed in experiment.

In case that the sample is covered by an adsorbate layer of thickness $d_{ads.}$ the angular intensity distributions for $Si^0$-$2p$ and $Si^{4+}$-$2p$ are additionally attenuated by adsorbate-specific damping factors, i.e., Eq. (A6) and Eq. (A7) then read as

$$I_{ads.}(Si^0 2p) \ = \ eq.(SI6) \cdot exp\left(-\frac{d_{ads.}}{\lambda_{Si^0 2p}(Ads.)\cdot cos\vartheta}\right) \qquad \text{(A13)}$$

$$I_{ads.}(Si^{4+} 2p) \ = \ eq.(SI7) \cdot exp\left(-\frac{d_{ads.}}{\lambda_{Si^{4+} 2p}(Ads.)\cdot cos\vartheta}\right) \qquad \text{(A14)}$$

If the adsorbates are simulated by a graphite layer the electron means free paths in the adsorbate of $Si^0$-$2p$ and the $Si^{4+}$-$2p$ photoelectrons are [36,37]

$$\lambda_{Si^0 2p}(Ads.) = 3.439\ nm \qquad \text{(A15)}$$

$$\lambda_{Si^{4+} 2p}(Ads.) = 3.432\ nm \qquad \text{(A16)}$$